\documentstyle[12pt,aps,prc,epsfig,preprint]{revtex}
\tightenlines
\begin{document}
\begin{titlepage}

\title{\bf Widths of $\bar K$-nuclear deeply
bound states in a dynamical model}

\author{J.~Mare\v{s} $^a$,   E.~Friedman $^b$,   A.~Gal $^b$ \\
$^a${\it Nuclear Physics Institute, 25068 \v{R}e\v{z}, Czech Republic} \\
$^b${\it Racah Institute of Physics, The Hebrew University,
Jerusalem 91904, Israel}}

\maketitle
\begin{abstract}

The relativistic mean field (RMF) model is applied to a system of
nucleons and a $\bar K$ meson, interacting via scalar and vector
boson fields. The model incorporates the standard RMF phenomenology
for bound nucleons and, for the $\bar K$ meson, it relates to low-energy
${\bar K}N$ and $K^-$ atom phenomenology. Deeply bound $\bar K$ nuclear
states are generated dynamically across the periodic table and are
exhibited for $^{12}$C and $^{16}$O over a wide range of binding
energies. Substantial polarization of the core nucleus is found for these
light nuclei. Absorption modes are also included dynamically,
considering explicitly both the resulting compressed nuclear density and
the reduced phase space for $\bar K$ absorption from deeply bound states.
The behavior of the calculated width as function of the $\bar K$ binding
energy is studied in order to explore limits on the possible existence
of narrow $\bar K$ nuclear states.
\newline
\newline
$PACS$: 13.75.Jz, 25.80.Nv, 36.10.Gv
\newline
\newline
Corresponding author: Avraham Gal, avragal@vms.huji.ac.il
\newline
tel: +972 2 658 4930, fax: +972 2 561 1519
\end{abstract}
\centerline{\today}
\end{titlepage}

\section{Introduction}
\label{sec:int}

The $\bar K N$ interaction near threshold is strongly attractive,
in agreement with the existence of the unstable bound state
$\Lambda (1405)$ below the $K^- p$ threshold.
The $\bar K$-nucleus interaction is also strongly attractive,
as derived from the strong-interaction shifts and widths in
kaonic-atom levels across the periodic table \cite{BFG97}.
It is not established yet how strong the $\bar K$-nucleus
potential is: `deep' (150-200 MeV \cite{FGB93,FGM99}) or
relatively `shallow' (50-60 MeV \cite{BGN00,CFG01})?
Is it possible to bind {\it strongly} $\bar K$ mesons in nuclei
and are such potentially deep bound states sufficiently narrow
to allow observation and identification? These issues have received
considerable phenomenological and theoretical attention recently
\cite{BGN00,CFG01,Kis99,AYa02,DHA04},
and some experimental evidence for candidate states in the
$(K^{-}_{\rm stop},n)$ and $(K^{-}_{\rm stop},p)$ reactions on $^4$He
(KEK-PS E471, \cite{ISB03,SBF04} respectively)
and in the $(K^{-},n)$ in-flight reaction on $^{16}$O
(BNL-AGS, parasite E930 \cite{KHA03}) has been presented very recently.
New experiments have been approved at KEK, using $(K^-,N)$ reactions to
search for $\bar K$ nuclear bound states \cite{Kis03}.
$(K^-,\pi^-)$ reactions \cite{YAk02} were also suggested in this context.

A prime concern in searching for $\bar K$ nuclear bound states is
the anticipated large width due to pionic conversion modes on
a single nucleon:
\begin{equation}
\label{eq:conv1}
{\bar K}N\rightarrow \pi\Sigma,\;\; \pi\Lambda \;\;\;\; (\sim 80\%)\;\;,
\end{equation}
with thresholds about 100 MeV and 180 MeV, respectively, below the
${\bar K} N$ total mass, and also due to non-pionic multi-nucleon
absorption modes, say
\begin{equation}
\label{eq:conv2}
{\bar K} N N \rightarrow Y N \;\;\;\; (\sim 20\%)\;\;,
\end{equation}
with thresholds about $m_{\pi}=140$ MeV lower than the single-nucleon
thresholds. The branching ratios in parentheses are
known from bubble-chamber experiments \cite{VVW77}.

The aim of the present work is to study {\it dynamical} effects
for $\bar K$ nuclear states in the range of binding
energy $B_{K} \sim 100 - 200$ MeV \cite{ISB03,SBF04,KHA03}
and in particular the width anticipated for such deeply bound states.
The relatively shallow chirally-motivated $\bar K$-nucleus potentials
\cite{BGN00,CFG01} which followed the microscopic construction by Ramos
and Oset \cite{ROs00} are of no use in this context, since they cannot
yield binding energy greater than the potential depth of about 50 MeV.
One must therefore depart from the microscopic approach in favor
of a more phenomenologically inclined model which is constrained
by data other than two-body $\bar K N$ observables.
The theoretical framework here adopted is the relativistic mean field
(RMF) model for a system of nucleons and one $\bar K$ meson
interacting through the exchange of scalar ($\sigma$) and vector
($\omega$) boson fields which are treated in the mean-field
approximation. By allowing the $\bar K$ to polarize the nucleons,
and vice versa, this dynamical calculation is made self consistent.
$\bar K$ absorption modes are included within a $t\rho$ optical-model
approach, where the density $\rho$ plays a dynamical role, and the
constant $t$ which is constrained near threshold by $K^-$-atom data
follows the phase-space reduction in reactions
Eqs. (\ref{eq:conv1},\ref{eq:conv2}) for a deeply bound $\bar K$.
A wide range of binding energies may be explored in the RMF calculation
simply by scanning over the coupling constants of the $\bar K$ meson
to the $\sigma$ and $\omega$ boson fields. Detailed calculations were
done by us across the periodic table. In this Letter we demonstrate the
essential points and conclusions for $^{12}$C and $^{16}$O
where the dynamical polarization effects are extremely
important for the energies of the $\bar K$ bound states as well as
for their widths. Akaishi et al. \cite{AYa02,DHA04} too
have found large polarization effects in lighter nuclei using
few-body variational techniques. The RMF is a systematic approach
used across the periodic table beyond the very light elements explored
by other techniques, and it can be used also to study multi-$\bar K$
configurations and to explore the $\bar K$ condensation limit
\cite{SGM94,SMi96}. Similar RMF calculations have been recently
reported for $\bar N$ states in nuclei \cite{BMS02,MSB04}.

\section{Methodology}
\label{sec:meth}

In the calculations described below, the standard RMF Lagrangian
$\cal L_N$ with the linear (L) parameterization of Horowitz and Serot
\cite{HSe81} as well as the nonlinear (NL) parameterization due to Sharma
et al. \cite{SNR93} are used for the description of the nucleonic
sector. The (anti)kaon interaction with the nuclear medium is
incorporated by adding to $\cal L_N$ the Lagrangian density
$\cal L_K$ \cite{SGM94,SMi96}:
\begin{equation}
\label{eq:Lk}
{\cal L}_{K} = {\cal D}_{\mu}^*{\bar K}{\cal D}^{\mu}K -
m^2_K {\bar K}K
- g_{\sigma K}m_K\sigma {\bar K}K\; .
\end{equation}
The covariant derivative
${\cal D_\mu}=\partial_\mu + ig_{\omega K}{\omega}_{\mu}$ describes
the coupling of the (anti)kaon to the vector meson $\omega$.
The vector field $\omega$ is then associated with a conserved current.
The coupling of the (anti)kaon to the isovector $\rho$ meson is here
excluded due to considering $N=Z$ nuclear cores in this initial report.

Whereas adding ${\cal L}_K$ to the original Lagrangian $\cal L_N$ does
not affect the form of the corresponding Dirac equation for nucleons,
the presence of $\bar K$ leads to additional source terms in the
equations of motion for the meson fields $\sigma$ and $\omega_0$ to
which the $\bar K$ couples:
\begin{eqnarray}
\left(-\Delta + m_{\sigma}^{2} \right) \sigma \; &=&
- g_{\sigma N}\rho_{S}
- g_{\sigma K} m_K {\bar K}K
+ \left( - g_{2}\, \sigma^{2} -
g_{3}\, \sigma^{3} \right) \;,
\label{eq:sigma}
\end{eqnarray}
\begin{eqnarray}
\left(-\Delta + m_{\omega}^{2} \right) \omega_0 &=&
+ g_{\omega N}\rho_{V}
- 2 g_{\omega K} (\omega_K+g_{\omega K}\omega_0) {\bar K}K \; ,
\label{eq:omega}
\end{eqnarray}
where $\omega_K$ is the $\bar K$ energy in the nuclear medium:
$$
\omega_K = \sqrt{m^2_K + g_{\sigma K} m_K \sigma + p^2_K}
- g_{\omega K}\omega_0\; ,
$$
and $\rho_{S}$ and $\rho_{V}$ denote the nuclear scalar and vector
densities, respectively. Adding $\bar K$ to the nuclear system
affects the scalar and vector potentials which enter the Dirac
equation for nucleons. This leads to the rearrangement,
or polarization of the nuclear core in the presence of $\bar K$.

In order to preserve the connection to previous studies
of kaonic atoms, the Klein Gordon (KG) equation of motion for the
$\bar K$ is written in the form \cite{CFG01}:
\begin{equation}
\label{eq:KG1}
\left[\Delta - 2{\mu}(B+V_{\rm opt}+V_c) + (V_c+B)^2 \right]{\bar K} = 0~~ ~~
(\hbar = c = 1).
\end{equation}
Here, $V_c$ denotes the static Coulomb potential for the $K^-$,
$\mu$ is the $\bar K$-nucleus reduced mass and
$B=B_K+{\rm i}{\Gamma_K}/2$ is the complex binding
energy. The real part of the $\bar K$ optical potential $V_{\rm opt}$
is then given by
\begin{equation}
\label{eq:VOP1}
{\rm Re}\;V_{\rm opt}={{m_K}\over{\mu}}({1\over{2}}S - V - {{V^2}\over{2m_K}})
\;\;\; ,
\end{equation}
where $S = g_{\sigma K}\sigma$ and $V = g_{\omega K}\omega_0$ are the
scalar and vector potentials due to the $\sigma$ and $\omega$ mean fields,
respectively.

Since the RMF approach does not address the imaginary part
of the potential, ${\rm Im}~V_{\rm opt}$ was taken in a phenomenological
$t\rho$ form, where its depth was fitted to the $K^-$ atomic data
\cite{FGM99}. Note that $\rho$ in the present calculations is no longer
a static nuclear density, but is a {\it dynamical} entity affected by the
$\bar K$ interacting with the nucleons via boson fields. The resulting
compressed nuclear density leads to increased widths, particularly for
deeply bound states. On the other hand, the phase space available for
the decay products is reduced for deeply bound states, which will act
to decrease the calculated widths. Thus, suppression factors multiplying
${\rm Im}~V_{\rm opt}$ were introduced from phase-space considerations,
taking into account the binding energy of the kaon for the initial decaying
state, and assuming two-body final-state kinematics for the decay products.
Two absorption channels were considered. In the first, Eq. (\ref{eq:conv1}),
a ${\bar K}N$ initial state decays into a $\pi Y$ final state.
The corresponding density-independent suppression factor is given by
\begin{equation}
\label{eq:spf1}
f_1=\frac{M_{01}^3}{M_1^3}\sqrt{\frac{[M_1^2-(m_\pi +m_Y)^2]
[M_1^2-(m_Y-m_\pi)^2]}
{[M_{01}^2-(m_\pi +m_Y)^2][M_{01}^2-(m_Y-m_\pi)^2]}}~
\Theta (M_1-m_{\pi}-m_Y)  \;\;\; ,
\end{equation}
\noindent
where $M_{01}~=~m_{\bar K}~+~m_N,~~M_1~=~M_{01}~-B_K$.
In the second absorption channel, Eq. (\ref{eq:conv2}),
a ${\bar K}NN$ initial state decays into a $YN$ final state.
The corresponding suppression factor is given by
\begin{equation}
\label{eq:spf2}
f_2=\frac{M_{02}^3}{M_2^3}\sqrt{\frac{[M_2^2-(m_N +m_Y)^2]
[M_2^2-(m_Y-m_N)^2]}
{[M_{02}^2-(m_N +m_Y)^2][M_{02}^2-(m_Y-m_N)^2]}}~
\Theta (M_2-m_Y-m_N)  \;\;\; ,
\end{equation}
\noindent
where $M_{02}~=~m_{\bar K}~+~2~m_N,~~M_2~=~M_{02}~-B_K$.
Although multi-nucleon absorption modes are often modeled to have
a power-law ${\rho}^{\alpha}$ ($\alpha$ $>$ 1) density dependence,
our comprehensive $K^-$-atom fits \cite{FGB93,FGM99} are satisfied
with $\alpha \sim 1$. We therefore assume $\alpha = 1$ in this
exploratory work also for this
second absorption channel,\footnote{a lucid theoretical discussion
of this point, with allowance for $\alpha = 1$, is due to Koltun
\cite{Kol79}} which means that $f_2$ too is independent of density.
We comment below on the effect of a possible density dependence of
$f_2$, reflecting perhaps a ${\rho}^2$ dependence of the non-pionic
decay mode (\ref{eq:conv2}) at high densities.

Since $\Sigma$ final states dominate these channels \cite{VVW77}
the hyperon $Y$ was here taken as $Y=\Sigma$.
Allowing $\Lambda$ hyperons would foremost {\it add} conversion
width to $\bar K$ states bound in the region $B_{K} \sim 100 - 180$ MeV.
For the combined suppression factor we assumed a mixture of 80\%
mesonic decay and 20\% nonmesonic decay \cite{VVW77}, i.e.
\begin{equation}
\label{eq:spf}
f=0.8~f_1~+~0.2~f_2 \; .
\end{equation}
This suppression factor is plotted as function of $B_K$ in the upper
part of Fig. \ref{fig:Gamma}, where a residual value of $f=0.02$,
when both $f_1$ and $f_2$ vanish, was assumed.

The coupled system of equations for nucleons and for the electromagnetic
vector field $A_0$, and for the mean fields $\sigma$ and $\omega_0$,
Eqs. (\ref{eq:sigma},\ref{eq:omega}) above, as well as the KG equation
(\ref{eq:KG1}) for $K^-$ were solved self-consistently using an
iterative procedure.
Obviously, the requirement of self-consistency is crucial for the proper
evaluation of the dynamical effects of the $\bar K$ on the nuclear
core and vice versa. We note that self-consistency is not imposed
here on the final-state hadrons which only enter through their
{\it on-shell} masses used in the phase-space suppression factors
given above. For the main $\pi \Sigma$ decay channel it is likely
that the attraction provided by the pion within a dynamical calculation
\cite{ROs00} is largely cancelled by the nuclear repulsion
deduced phenomenologically for $\Sigma$ hyperons \cite{MFG95,NSA02,SNA04}.

\section{Results}
\label{sec:res}

The main objective of the present calculations of $\bar K$-nucleus
bound states was to establish correlations between various observables
such as the $\bar K$ binding energy, width and macroscopic nuclear
properties. In particular we aimed at covering a wide range of binding
energies in order to evaluate widths of possible strongly bound $\bar K$
states. Furthermore, in order to study effects of the nuclear polarization,
we calculated rms radii and average densities of the nuclei involved and,
in some cases, also single particle energies.
Extensive calculations were made for $^{12}$C and $^{16}$O, nuclei
that had been discussed earlier in the context of strongly bound $\bar K$
states \cite{Kis99,KHA03}. Additionally, more restricted calculations were
made for $^{40}$Ca and $^{208}$Pb.

In a preliminary test we performed dynamical calculations for $K^-$
{\it atomic} $1s$ states, which produced only negligibly small
polarization effects, thus validating previous analyses of kaonic atom data.
The empirical values $g^{(1)}_{\sigma K}$ and $g^{(1)}_{\omega K}$,
as found from a fit to kaonic atom data \cite{FGM99},
were therefore used as a starting point for calculations.
A full dynamical calculation was then made for $K^-$ {\it nuclear} states
starting from the $t\rho$ imaginary potential obtained from the atomic fit,
while entering dynamically in the iteration cycles the resulting nuclear
density and the suppression factor $f$ as defined by Eq.(\ref{eq:spf}).
This {\it dynamical} calculation nearly doubled, for the light nuclei
here considered, the depth of the real part of the phenomenological
{\it static} $K^-$-nucleus potential of Ref. \cite{FGM99} which is of
the `deep' variety, typically 150-200 MeV deep in the static calculation.

Since there is no preferred way of varying the depth of the real
$K^-$-nucleus potential in order to produce different values of binding
energies, we used two methods for scanning over binding energies.
The first one, referred to below as the `RMF' method, was to scale down
successively $g_{\sigma K}$ from its initial value $g^{(1)}_{\sigma K}$
and, once it reached zero, to scale down $g_{\omega K}$ too from its
initial value $g^{(1)}_{\omega K}$ until the $K^-$ $1s$ state became
unbound. As an alternative method we chose as a starting point values
$g^{(2)}_{\sigma K}$ and $g^{(2)}_{\omega K}$ obtained from fits to
kaonic atom data where the RMF-based real potential was joined at large
radii by a $`t \rho $' expression, using for $t$ the chiral ${\bar K}N$
amplitudes of Ramos and Oset \cite{ROs00}. This starting potential
was of the `shallow' variety \cite{CFG01}, about 55 MeV deep in the static
calculation. It is gratifying that the replacement of the
chiral $`t \rho $' expression within the nucleus by the RMF model did not
change the resulting $\bar K$-nuclear potential depth. We then
increased the potential depth by scaling up $g_{\sigma K}$ from its initial
value $g^{(2)}_{\sigma K}$, in order to achieve as deep binding as in
the first set of calculations, while keeping $g_{\omega K}$ constant at
its initial value $g^{(2)}_{\omega K}$. This second method will be referred
to below as the `chiral tail' method. Note that in both methods ($j=1,2$)
the starting values ($g^{(j)}_{\omega K}, g^{(j)}_{\sigma K}$) correspond
to potentials that produce in the dynamical calculation good fits to the
$K^-$ atomic data, although Re $V_{\rm opt}^{(j)}$ have vastly different
depths. The good fits to the atomic data are inevitably lost once the
coupling constants ($g_{\sigma K}$ in the procedure outlined above) are
allowed to vary in order to scan over a wide range of binding energies
and the associated widths.

\begin{figure}
\epsfig{file=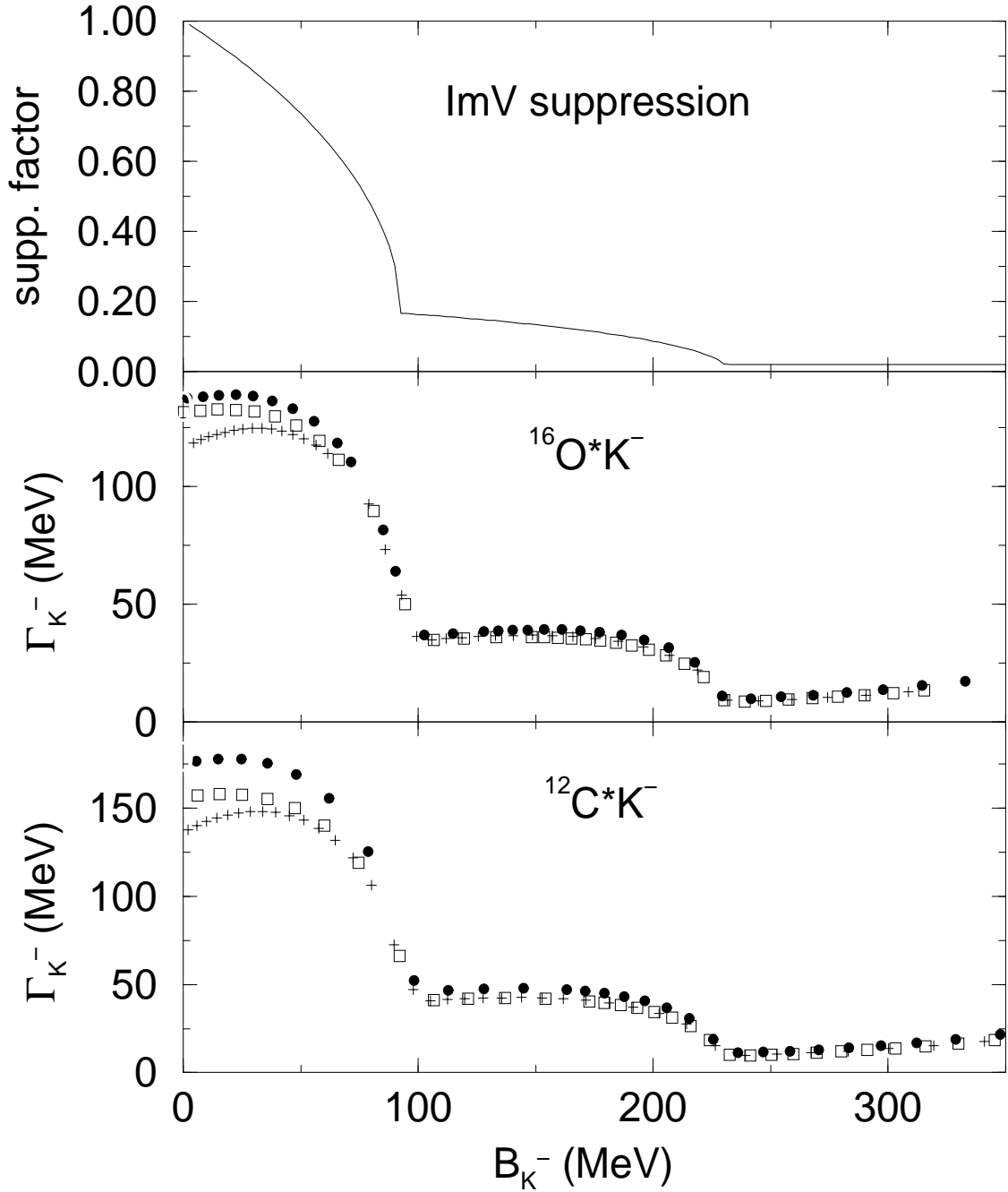, height=180mm,width=150mm}
\caption{Phase-space suppression factor for the imaginary potential (top),
and widths of the $1s$ $K^-$-nuclear state (middle: in $^{16}$O,
bottom: in $^{12}$C) as function of the $K^-$ binding energy
(see text for symbols).}
\label{fig:Gamma}
\end{figure}

Figure \ref{fig:Gamma} shows, in its middle and lower parts, calculated
widths $\Gamma_{K^-}$ as function of the binding energy $B_{K^-}$ for $1s$
states in $^{~~16}_{K^-}$O and $^{~~12}_{K^-}$C, respectively.
Open squares and solid circles are for the L and NL versions of
the RMF model, respectively, both
calculated using the `RMF' method for scanning the binding energies.
The crosses are for RMF-L, using the `chiral tail' method for scanning the
binding energies. It is clearly seen that the widths of the $K^-$
nuclear state follow closely the dependence of the suppression factor on
the binding energy and that, except for binding energies smaller than
50 MeV, the dependence of the width on the binding energy follows,
for a given nucleus, almost a universal curve. We note that the widths
calculated in the range $B_K \sim 100 - 200$ MeV assume values
$40 \pm 5$ MeV, which are considerably larger than what the suppression
factor of the upper part of the figure would suggest.
This is largely related to the dynamical nature of the RMF
calculation whereby the nuclear density is increased by the
polarization effect of the $K^-$, as shown in the next figures.
Furthermore, replacing $\rho$ by ${\rho}^2$ for the density dependence of
the non-pionic decay modes (\ref{eq:conv2}) is estimated to increase the
above values of the width by 10 - 15 MeV. This estimate follows, again,
from the increase of nuclear density with $B_K$ noticed above.
Switching on the $\pi \Lambda$ decay mode would increase further this
estimate by 5-10 MeV in the range $B_K \sim 100 - 180$ MeV.
We assert that the estimate $\Gamma _{K} = 40 \pm 5$ MeV in the range
$B_K \sim 100 - 200$ MeV provides a reasonable lower bound on the width
expected in any realistic calculation. A more detailed systematics is
defered to a subsequent regular report.

\begin{figure}
\epsfig{file=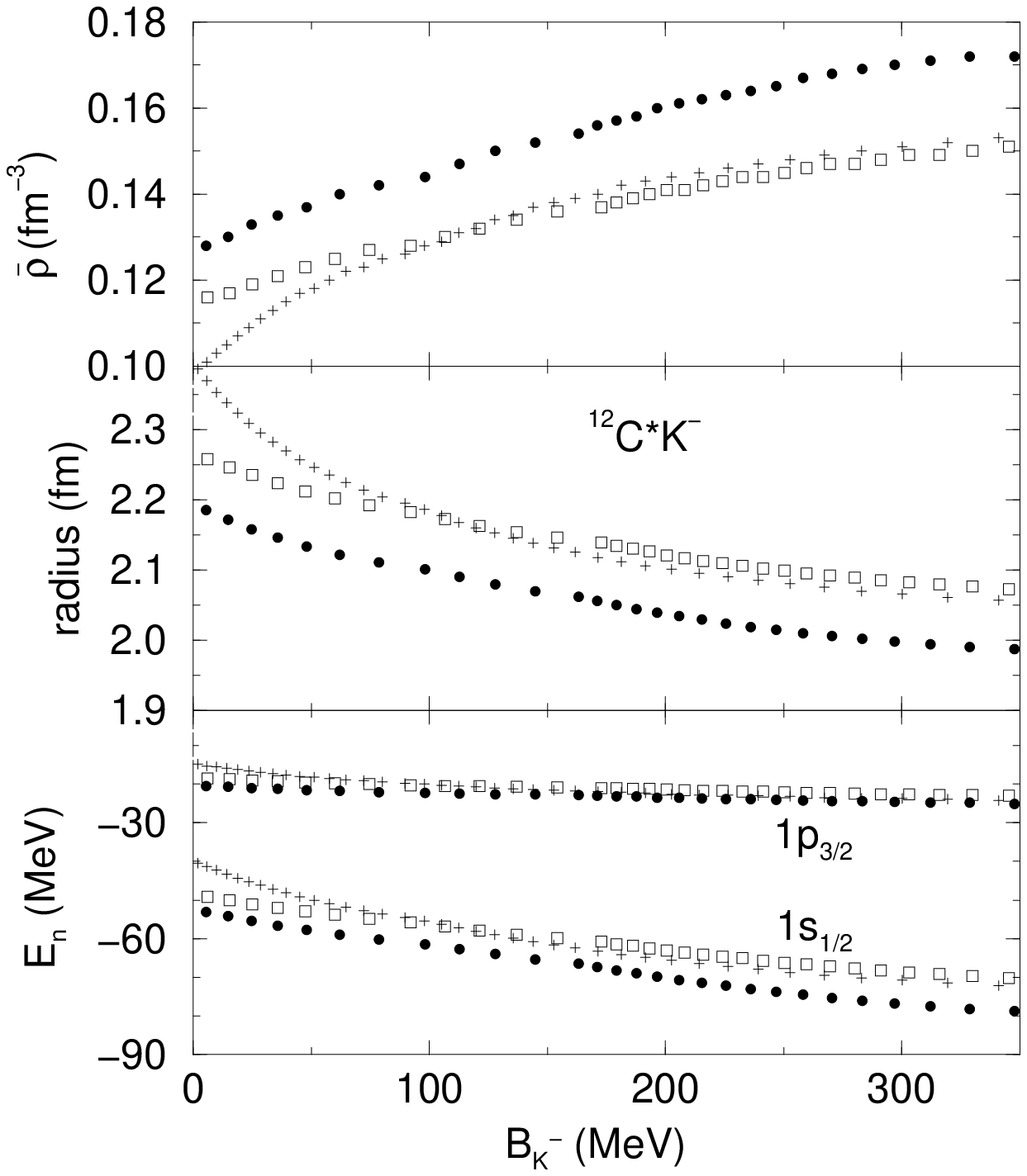, height=180mm,width=150mm}
\caption{Average nuclear density, nuclear rms radius and neutron
single-particle energies for $^{~~12}_{K^-}$C. Symbols are as in
Fig. \ref{fig:Gamma}.}
\label{fig:nuclC}
\end{figure}

\begin{figure}
\epsfig{file=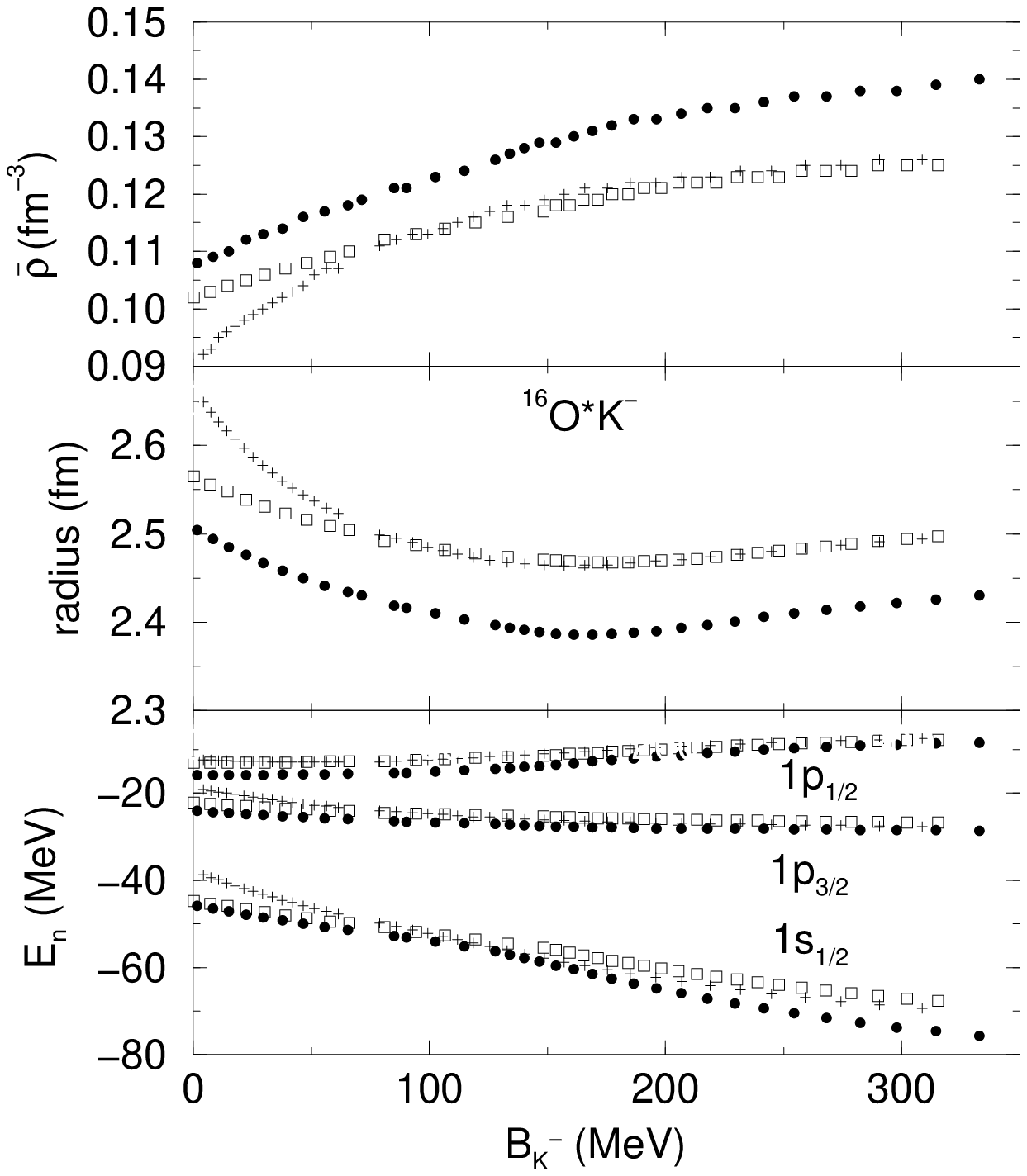, height=180mm,width=150mm}
\caption{Average nuclear density, nuclear rms radius and neutron
single-particle energies for $^{~~16}_{K^-}$O. Symbols are as in
Fig. \ref{fig:Gamma}.}
\label{fig:nuclO}
\end{figure}

Figures \ref{fig:nuclC} and \ref{fig:nuclO} exhibit various nuclear
properties for $1s$ states in $^{~~12}_{K^-}$C and $^{~~16}_{K^-}$O,
respectively.
The top and middle parts show the calculated average nuclear density
$\bar \rho = \frac{1}{A}\int\rho^2d{\bf r}$ and the nuclear rms radius,
respectively, and the lower parts show the $1s$ and $1p$ neutron
single-particle energies $E_n$.
The differences between the linear and non-linear models reflect the
different nuclear compressibility and the somewhat different nuclear
sizes obtained in the two models. Again, for $K^-$ binding energy greater
than 50 MeV the results are independent of the way the binding energy is
being scanned. It is interesting to note that the increase in the
nuclear rms radius of $^{~~16}_{K^-}$O for large values of $B_{K^-}$ is
the result of the reduced binding energy of the $1p_{1/2}$ state,
due to the increased spin-orbit term. Note also that as $B_{K^-}$
approaches zero we do {\it not} recover the values inherent in static
calculations for the various nuclear entities because the coupling
constants $g_{\sigma K}$ and $g_{\omega K}$, and the imaginary part of
the potential, still assume nonzero values. The substantial increase of
$\bar \rho$, the decrease of the nuclear rms radius,
and the decrease of the $1s$ neutron single-particle energy, all
point out to a significant polarization of the nuclear core by the $1s$
$K^-$. Finally, we note that in
similar calculations for $^{40}$Ca and $^{208}$Pb the widths of the
$K^-$ nuclear state {\it vs.} its binding energy turned out to be
similar in shape to the corresponding results for $^{12}$C and $^{16}$O,
but the effects on the average density and the rms radius are negligibly
small, as expected for heavier nuclei.

Table \ref{tab:tabl1} shows several examples of the excited-state spectrum
of $K^-$-nuclear states in the core of $^{16}$O. The calculated widths
of the excited states appear to follow the general trend of widths
for $1s$ states as shown in Fig. \ref{fig:Gamma}. The polarization
of the nucleus as judged by the value of its average density
$\bar {\rho}$ appears to diminish, the higher the $K^-$ state is,
in agreement with Fig. \ref{fig:nuclO}. We note the fairly large
spacing between neighboring states: the effective $\hbar \omega$
is of order 100 MeV for the first spectrum, decreasing to 80 MeV and
to 70 MeV in the second and third spectra, respectively, as the $1s$
binding is made lower. Static calculations give smaller spacings.
Kishimoto et al. \cite{KHA03} have very recently suggested evidence
from the measured neutron spectrum in the $^{16}{\rm O}(K^-,n)$ reaction
for a peak at $B_K \sim 90$ MeV which they interpreted as the
$1p$ $\bar K$ state, probably since a hint for a peak at
$B_K \sim 130$ MeV is also suggested by the same spectrum.
The results shown in Table \ref{tab:tabl1} rule out such interpretation
which would require $\hbar \omega \sim 40$ MeV, considerably less than
our {\it dynamical} calculations produce. If the peak at
$B_K \sim 90$ MeV were a $1p$ state, one should have expected the $1s$ state
to be deeply bound, at $B_K \sim 200$ MeV where no signal has been observed.
On the other hand, if this peak is a $1s$ state, then the relatively
shallow $1p$ state would be expected too broad to be distinguished in the
data.

\section{Conclusions}
\label{sec:conc}

Binding energies and widths of deeply bound $K^-$ nuclear states in
$^{12}$C and $^{16}$O were calculated with the aim of establishing
values of widths that could be expected for binding energies in the
range of 100-200 MeV. The method chosen was to couple the $K^-$
dynamically to the nucleus within the RMF approach, which is applicable
throughout the periodic table, but which is disconnected from near-threshold
${\bar K}N$ phenomenology. Negligible polarization effects were found
for {\it atomic} states, which confirms the optical-potential phenomenology
of kaonic atoms as a valid starting point for the present study.
Substantial polarization of the core nucleus was found in these light nuclei
for deeply bound $\bar K$ nuclear states.
Almost universal dependence of antikaon widths on its binding energy was
found, for a given nucleus, suggesting that the details of how the calculated
binding energy is varied over the desired range of values is largely
immaterial. The widths are mostly determined by the phase-space suppression
factors on top of the increase provided by the density of the compressed
nuclei. The present results already provide useful
guidance for the interpretation of recent experimental results \cite{KHA03}
by placing a lower limit $\Gamma_K \sim 35 - 40$ MeV on $\bar K$ states
in $^{16}$O bound in the range $B_K \sim 100 - 200$ MeV.
For lighter nuclear targets such as $^4$He, where the RMF approach becomes
unreliable but where nuclear polarization effects are found larger using
few-body calculational methods \cite{AYa02,DHA04}, we anticipate larger
widths for $\bar K$ deeply bound states, if such states do exist \cite{SBF04}.
\newline
\newline
This work was supported in part by the GA AVCR grant IAA1048305
and by the Israel Science Foundation grant 131/01.

\begin{table}
\caption{$K^{-}$ bound-state spectra in $^{~~16}_{K^-}\rm O$ calculated for
several RMF (NL) Lagrangians specified by different coupling-constant
ratios $\alpha_{\sigma} = g_{\sigma K}/g^{(1)}_{\sigma K}$ and
$\alpha_{\omega} = g_{\omega K}/g^{(1)}_{\omega K}$.
The static average density for $^{16}\rm O$ is
$\bar {\rho} = 0.100$ fm$^{-3}$.}
\label{tab:tabl1}
\begin{center}
\begin{tabular}{llcccc}
$\alpha_{\sigma}$ & $\alpha_{\omega}$ & $nl$ & $B_{K^-} ~\rm {(MeV)}$ &
$\Gamma_{K^-} ~\rm {(MeV)}$ & $\bar {\rho} ~(\rm {fm}^{-3})$ \\
\hline
0.45 & 1 & $1s$ & 196.1 & 35.0 & 0.133 \\
& & $1p$ & 82.2 & 83.0 & 0.127 \\
& & $2s$ & 3.7 & 89.9 & 0.111 \\
\hline
0.05 & 1 & $1s$ & 133.9 & 38.7 & 0.127 \\
& & $1p$ & 50.6 & 119.0 & 0.120 \\
\hline
0 & 0.85 & $1s$ & 90.2 & 64.2 & 0.121 \\
& & $1p$ & 23.8 & 124.5 & 0.115 \\
\end{tabular}
\end{center}
\end{table}

\end{document}